\documentclass[reprint, amsmath,amssymb, aps]{revtex4-2}

\usepackage{graphicx}
\usepackage{dcolumn}
\usepackage{bm}
\usepackage[version=4]{mhchem}
\usepackage[detect-display-math=true,per-mode=symbol]{siunitx} 

\newcommand{\effcond}{\sigma}
\newcommand{\condmap}{g}
\newcommand{\condnet}{g}
\newcommand{\bondcond}{g}

\begin{document}

\preprint{APS/123-QED}

\title{Percolation and conductivity in evolving disordered media}

\author{Carl Fredrik Berg}
\email{carl.f.berg@ntnu.no}
\affiliation{PoreLab, Department of Geoscience and Petroleum, Norwegian University of Science and Technology, Trondheim, Norway}%

\author{Muhammad Sahimi}%
 
\affiliation{Mork Family Department of Chemical Engineering and Materials 
Science, University of Southern California, Los Angeles, California 
90089-1211}%

\date{\today}

\begin{abstract}
Percolation theory and the associated conductance networks have provided deep insights into the flow and transport properties of a vast number of heterogeneous materials and media. In practically all cases, however, the conductance of the networks' bonds remains constant throughout the entire process. There are, however, many important problems in which the conductance of the bonds evolves over time and does not remain constant. Examples include clogging, dissolution and precipitation, catalytic processes in porous materials, as well as the deformation of a porous medium by applying an external pressure or stress to it that reduces the size of its pores. We introduce two percolation models to study the evolution of the conductivity of such networks. The two models are related to natural and industrial processes involving clogging, precipitation, and dissolution processes in porous media and materials. The effective conductivity of the models is shown to follow known power laws near the percolation threshold, despite radically different behavior both away from and even close to the percolation threshold. The behavior of the networks close to the percolation threshold is described by critical exponents, yielding bounds for traditional percolation exponents. We show that one of the two models belongs to the traditional universality class of percolation conductivity, while the second model yields non-universal scaling exponents.
\end{abstract}

\maketitle


\section{\label{sec:intro}Introduction}

Percolation theory \cite{stauffer_introduction_2003,saberi_recent_2015} has 
provided deep insights into the flow and transport properties of a vast number 
of heterogeneous materials and media, and has found numerous applications 
\cite{sahimi_applications_2023} in a variety of contexts. In many cases the 
heterogeneous materials are represented by conductance networks 
\cite{kirkpatrick_percolation_1973}, if a scalar transport process is to be 
studied; by a network of elastic elements, such as springs 
\cite{feng_percolation_1984-1,feng_percolation_1984,feng_position-space_1985} 
or beams \cite{lewinski_dynamical_1988}, if vector transport processes are 
investigated, or by a network of interconnected pores \cite{sahimi_flow_2011} 
if one is to examine various fluid flow phenomena in porous materials and 
media. When representing natural and industrial heterogeneous materials, the 
conductance of the bonds or pores might be distributed according to some
probability distribution function that represents the morphology of the 
materials \cite{balberg_recent_1987,balberg_continuum_2009}. In practically all
cases, however, the conductance of the network elements is modeled as constant 
throughout the percolation process under study.

There are, however, many important problems in which the conductance of the 
bonds in the networks that represent the morphology of the system of interest 
evolves over time and, therefore, does not remain constant. One example is 
non-catalytic gas-solid reactions with solid products, such as sulphation of 
calcined limestone particles that are highly porous and contain a range of 
pore sizes,
\begin{displaymath}
\mbox{CaO(s)+SO$_2$(g)+$\frac{1}{2}$O$_2$(g) $\to$ CaSO$_4$(s)}
\end{displaymath}
Numerous experiments indicate \cite{reyes_percolation_1987,shah_transport_1987}
that during the reaction the solid volume increases, and the pores are 
gradually plugged. Another example is the important problem of catalyst 
deactivation \cite{sahimi_percolation_1985} in which a reactant reacts within 
the pore space of the catalyst and produces products that not only cover the 
catalyst's active sites, but also precipitate on the solid surface of the pores
and plug them, leading to deactivation of the catalyst. A third example is the 
transport of colloidal particles and stable emulsions in flow through a porous 
medium, during which the particles and emulsions precipitate on the surface of 
the pores and reduce their flow capacity \cite{rege_network_1987,sahimi_hydrodynamics_1991,schwartz_particle_1993}. The pore space of rock and 
other natural porous media evolve due to dissolution or precipitation. The fourth example is quartz cementation in sandstone that yields a pore space with
a continuous range of various porosity and the corresponding flow and transport
properties, such as permeability and electrical conductance. Another example is
the evolution of sandstone pore structure in the near-well region by salt 
precipitation during CO$_2$ injection for its sequestration \cite{miri_salt_2016,jeddizahed_experimental_2016}, as well as during evaporation of brine and 
the resulting salt precipitation \cite{rad_pore-scale_2013,rad_effects_2015,dashtian_pore-network_2018}. Pore structure evolution is also observed in systems where the pore sizes of porous materials and, hence, their conductances, are reduced mechanically by, for example, applying an external stress or pressure to the material \cite{richesson_flow_2021,richesson_flow_2021-1}. In all such cases, and numerous other examples, 
such as clogging of nanopores by transport of DNA \cite{kubota_clog_2019}, one 
has an evolving network.

Thus, the purpose of the present paper is to study the transport properties of 
evolving networks, particularly near their percolation threshold $p_c$. The goal of our study is twofold. One is to understand how the transport properties
evolve in such networks, and how their evolution depends on the manner by which
the conductances decrease. The second goal is to see whether the power-law of 
percolation theory, according to which the effective conductivity $\effcond_e$ follows
the universal power law,
\begin{equation}
\effcond_e\propto (p-p_c)^t\;,
\label{eq:condPowerLaw}
\end{equation}
is also satisfied by the effective conductivity of evolving networks, where 
$p$ is the fraction of the bonds with a non-zero conductance, and $t$ is the 
critical exponent whose value is largely universal with $t\simeq 1.3$ in two 
dimensions.

The rest of this paper is organized as follows. In the next section, we 
introduce the models that we study and explain how they are employed in our
numerical simulations. In Section III, we present the details of the numerical 
simulations. Section IV presents the results for the power laws that the 
effective conductivity of the proposed models follow near the percolation 
threshold, and compares them to the traditional models of random conductance 
networks. In Sec. V the implications of the results are discussed in detail, 
while the last section summarizes the results.

\section{\label{sec:models}The models}

The main motivation for this work is transport in evolving porous media, which typically occurs in complex three-dimensional pore networks. For the sake of more efficient simulations of very large networks, however, we restrict our study to the square lattice, which allows us to make precise comparisons with the existing models. 

The simplest network we consider is the traditional square lattice in which we remove bonds by a probability $p$ and where the remaining bonds have unit conductance. For straightforward comparison with models 
that will be introduced below, we define these networks in the following way 
\cite{grimmett_percolation_1999}: We attribute a random number $p(e)\in[0,1]$ 
to each bond $e\in E$, where $E$ is the set of bonds in the initial graph, in our case the square lattice. This gives rise to a conductance map $\condmap_o \colon E \to \mathbb{R}^+$ by letting 
\begin{equation}
\condmap_o(p,e) = 
\begin{cases}
1 &\text{ if } p(e) \leq p \\
0 &\text{ if } p(e) > p
\end{cases}
\label{eq:evolveOne}
\end{equation}
Thus, we attribute unit conductance to all the bonds with a random number 
smaller than $p$, and zero conductance to the remaining bonds. A conductance map $\condmap$ defines a network, i.e., a weighted graph where the weights represent conductances. To simplify notation we let $\condmap$ represent both the conductance map and the network defined by this map. 
Networks $\condnet_o$ in which the bonds (or sites) are removed (make no contribution to transport) by a certain probability have been widely studied in the classical percolation theory, and are well covered in the literature \cite{stauffer_introduction_2003,saberi_recent_2015,sahimi_applications_2023}. They have many interesting properties with known behavior close to the percolation threshold $p_c$. 

In the model above all bonds have unit conductance. Different transport 
processes have different relations to, e.g., the cross-sectional area available
for transport. For example, the electrical conductance of a cylindrical pipe 
with a constant cross-sectional area and filled with an electrolyte is 
proportional to the cross-sectional area, whereas, according to the 
Hagen-Poiseuille equation, the fluid flow rate through the same cylindrical 
pipe due to a pressure difference is proportional to the cross-sectional area 
squared. If we view a bond as a cylindrical pipe of unit length and a variable 
volume $V_b$, then the cross-sectional area will be proportional to the volume,
$A_b\propto V_b$. If a bond weight is assumed to represent its volume or mass, 
then different transport processes can be represented by raising the weight to 
a power. In this article, we use mass, instead of volume. For a porous medium, 
this can be thought of as the mass of the electrolyte or a fluid filling the 
volume, thereby equating the two through a constant electrolyte of fluid 
density.

Motivated by evolving porous media, we introduce two types of evolving 
networks. The first is similar to the networks defined by Eq.~\eqref{eq:evolveOne}, but where we have a link weight that is inversely 
proportional to the probability that the bond is removed. The link weight is 
set to be equal to the mass, and is expressed as
\begin{equation}
m_p(p,e) = 
\begin{cases}
1 - p(e) &\text{ if } p(e) \leq p \\
0 &\text{ if } p(e) > p
\end{cases}
\label{eq:evolveOrg}
\end{equation}
This type of network is related to clogging of a porous medium, such as a filter or membrane. The model also has a close correspondence with the aforementioned non-catalytic gas-solid reactions and catalyst deactivation when diffusion limits the rate of reaction. As a result, the sizes of the pores are not reduced uniformly. In all such processes, the phenomena begin in a fully connected network, but, over time, the size of the pores gradually decreases due to either a chemical reaction that produces solid products (as in diffusion-limited catalytic or non-catalytic reactions), or by the precipitation of particles on surface of the pores due to the physical interactions between the particles and the pore surface, as in the clogging problems.

The initial network before the onset of the closure process has a mass distribution where $1-p(e)$ is the mass of bond $e$, and the blocking of a bond tends to happen at the least conductive bonds, i.e., the bonds with the 
smallest mass, thus the smallest $1-p(e)$ values. As discussed above, when the 
link weight is considered as a mass (or volume), then the weight can be related
to various types of transport processes through an exponent $\tau$ as, 
$\condmap_p^\tau(p,e)=m_p(p,e)^\tau$. As described above, $\tau=1$ is related to 
electrical conductance, while $\tau=2$ is related to fluid flow. For this type 
of network, the values of the bond conductance have constant value 
$(1-p(e))^\tau$ until removed depending on $p$. The conductance distribution 
for the network evolves, however, with $p$.

A third type of network is given by the following function:
\begin{equation}
m_s(p,e) = 
\begin{cases}
p - p(e) &\text{ if } p(e) \leq p \\
0 &\text{ if } p(e) > p
\end{cases}
\label{eq:evwwolveOrg}
\end{equation}
which is a simple representation of a precipitation/dissolution process, where 
the precipitation (or, equivalently, the dissolution) is similar throughout 
the network. This corresponds to the aforementioned catalytic or non-catalytic 
gas-solid in which diffusion plays no role, and only the kinetics of the 
reactions are important. For a porous medium, the precipitation reduces the 
volume of the pores, thereby reducing the original mass $1-p(e)$ by the same 
mass $1-p$ throughout the network, resulting in a mass of $(1-p(e))-(1-p)=
p-p(e)$. Once again, we relate the mass to transport through the exponent 
$\tau$ as $\condmap_s^\tau(p,e)=m_s(p,e)^\tau$. For this type of network, both 
the bond conductances and their distribution evolve with $p$.

For comparison to the evolving networks that we have introduced, we also 
consider more traditional networks with a uniform mass distribution between 
endpoints $a$ and $b$; $U(a,b)$, with $0\leq a<b\leq 1$. Each bond $e$ has two 
associated probabilities, one for the probability $p(e)\in [0,1]$ of being 
removed, and one for the mass $m(e)\in [a,b]$ being a random number between 
$a$ and $b$. The model is then defined by
\begin{equation}
m_r(a,b) = 
\begin{cases}
m(e) &\text{ if } p(e) \leq p \\
0 &\text{ if } p(e) > p
\end{cases}
\label{eq:nonEvolveRandom2}
\end{equation}
Here, we only keep the end-points from the distribution $U(a,b)$ in our 
notation. This mass model then gives rise to the conductance model 
$\condmap_r^\tau(a,b)=m_r(a,b)^\tau$, so that the mass distribution stays equal 
to $U(a,b)$ for all $p$ and, as a consequence, the conductance distribution 
does {\it not} evolve with $p$. Later in this paper, we will demonstrate that 
this type of network is similar to our evolving networks for a restricted 
range of $p$. As the properties of the $\condmap_r^\tau(a,b)$ models are known 
in the literature \cite{kogut_distribution-induced_1979,straley_non-universal_1982,feng_transport_1987}, they will be valuable for comparison with our 
evolving networks.

Note that the unit conductance in the $\condnet_0$ model means that we can 
equate the conductance map $\condmap_0^\tau(p,e)$ to a mass model $m_0(p,e)$ for 
all $\tau$. We drop the superscript $\tau$ for the $\condnet_0$ models, as they 
are all equal.

\section{Computer simulations}

All calculations in this study were carried out using the Python programming  
language. The networks were stored as two lists, one for the vertices, or 
sites, and one for the edges, or bonds. The reason for using lists instead of, 
e.g., NumPy arrays (a Python library), is that they are used in several loops, 
where retrieving values from lists is faster than from arrays. The vertex list 
stores for each vertex the coordinates, the number of edges connected to the 
vertex, and the edges identification numbers. The edge list stores the edge 
identity, the associated random number $p(e)\in [0,1]$, and the identification 
numbers for the two connected vertices.

Two opposite sides of the networks were considered as the inlet and outlet. For
each network, we first determine the percolation threshold $p_c$, i.e., the 
smallest value of $p$ such that the network $\condnet_o(p)$ connects the inlet 
to the outlet. The threshold was computed by a binary search algorithm: The 
links are ordered according to their value of $p(e)$. We start the binary 
search by checking if $\condnet_o(p)$ is connected when $p$ equals the link value 
$p(e)$ in the middle of the stack. If it is connected, we remove the upper half
of the link stack; if not, we remove the lower half. We then check if 
$\condnet_o(p)$ is connected for $p$ equal to the link value $p(e)$ in the middle
of the remaining stack. This process is continued until there is only one link 
left in the stack, yielding the bridging link at the percolating threshold. In 
addition, we check during the binary search whether the network is connected by
first performing two breadth-first searches \cite{tarjan_depth-first_1972,
sheppard_invasion_1999}, one from the inlet and one from the outlet, and then 
checking the intersection of the resulting two searches; the network is 
connected if the intersection is nonzero.

To calculate the effective conductance of the networks, we follow the standard 
approach, namely, applying Kirchhoff's circuit laws. For each node $i$
we have the equation
\begin{equation}
\sum_j \condmap(e)(\phi_j - \phi_i) = 0 \quad ,
\label{eq:kirchoff}
\end{equation}
where $\phi_i$ is the potential at node $i$, and $e$ is the edge $(i,j)$ for 
the set of nodes $\{j\}$ connected to $i$. The effective conductance is 
computed by representing the set of equations given by Eq.~\ref{eq:kirchoff} in
matrix form: ${\bf M \Phi}={\bf B}$ 
\cite{kirkpatrick_percolation_1973}. Here, {\bf B} is the vector representing 
the boundary conditions. As the boundary conditions, we applied a potential 
difference between the inlet and outlet. The matrix {\bf M} represents the 
discretized Laplacian matrix for the network with the conductance values as 
weights for the bonds, and is stored in compressed sparse column matrix format 
using the SciPy library. The matrix $\mathbf{M}$ was inverted using either the 
conjugate-gradient or the LU-decomposition method, both in the SciPy library,
depending on the bandwidth of the matrix. We then obtained the solution vector 
${\bf \Phi}={\bf M}^{-1}{\bf B}$, which yields the potentials $\phi_i$ in the 
nodes, from which the total current through the network and, hence, the 
effective conductance is computed. Dividing the effective conductance by the 
network size we obtain the effective conductivity $\effcond_e$ \cite{stauffer_introduction_2003,sahimi_applications_2023}.

For well-connected networks, the approach was efficient and accurate. Close to 
the percolation threshold, however, where, due to the tortuous and constricted 
nature of the conducting paths, the current is very unevenly distributed in 
the network, the matrix inversion is susceptible to numerical errors. To reduce
such numerical issues, we construct the Laplacian matrix {\bf M} of the 
backbone, where we identify the backbone of the network by a method similar to 
Tarjan's strongly connected components algorithm \cite{tarjan_depth-first_1972,
sheppard_invasion_1999}, but with a non-recursive implementation in order to 
avoid stack overflow problems for large network sizes. For each network size, 
we generated at least 100 realizations and averaged the results.

\section{Results and discussion}

We now investigate the evolving networks introduced in Section 
\ref{sec:models}, both theoretically and numerically. We carried out extensive 
simulations in order to observe and study the behavior of the effective 
conductivity of the networks as they evolve.

\subsection{Conductance functions $\condmap_o$ and $\condmap_p^\tau$}

\begin{figure*}[t]
\begin{center}
\includegraphics[width=2\columnwidth]{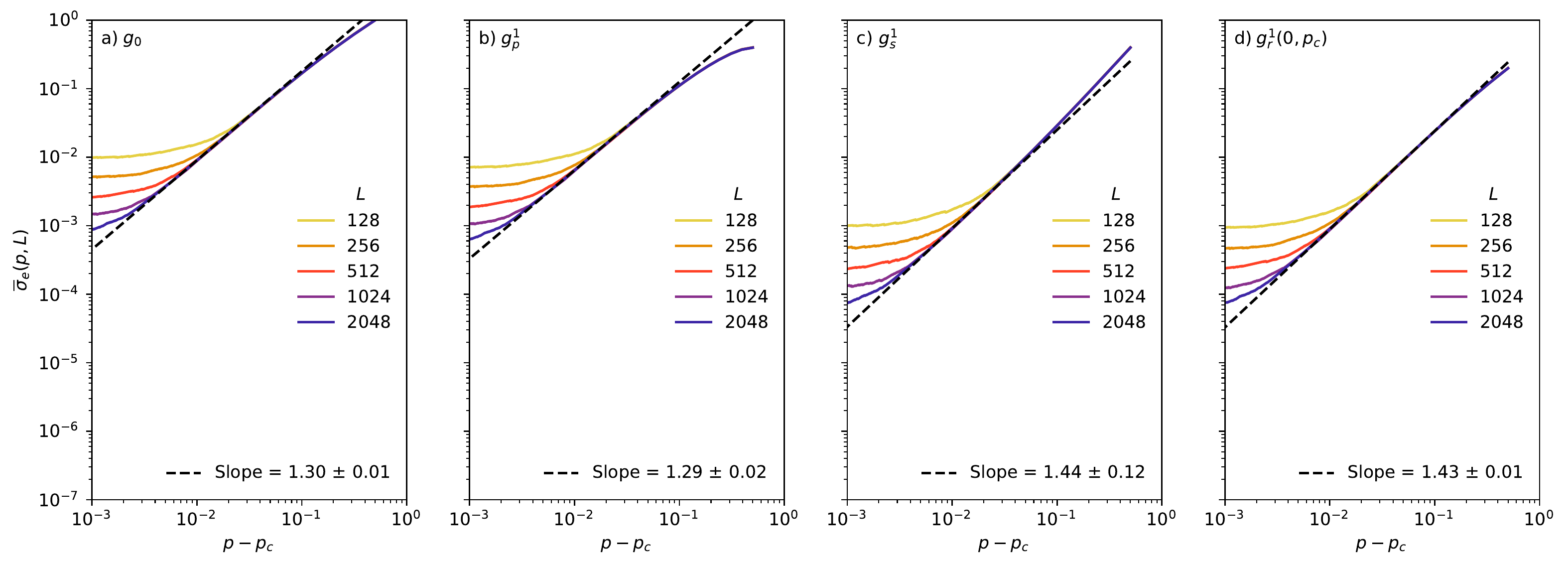}
\includegraphics[width=2\columnwidth]{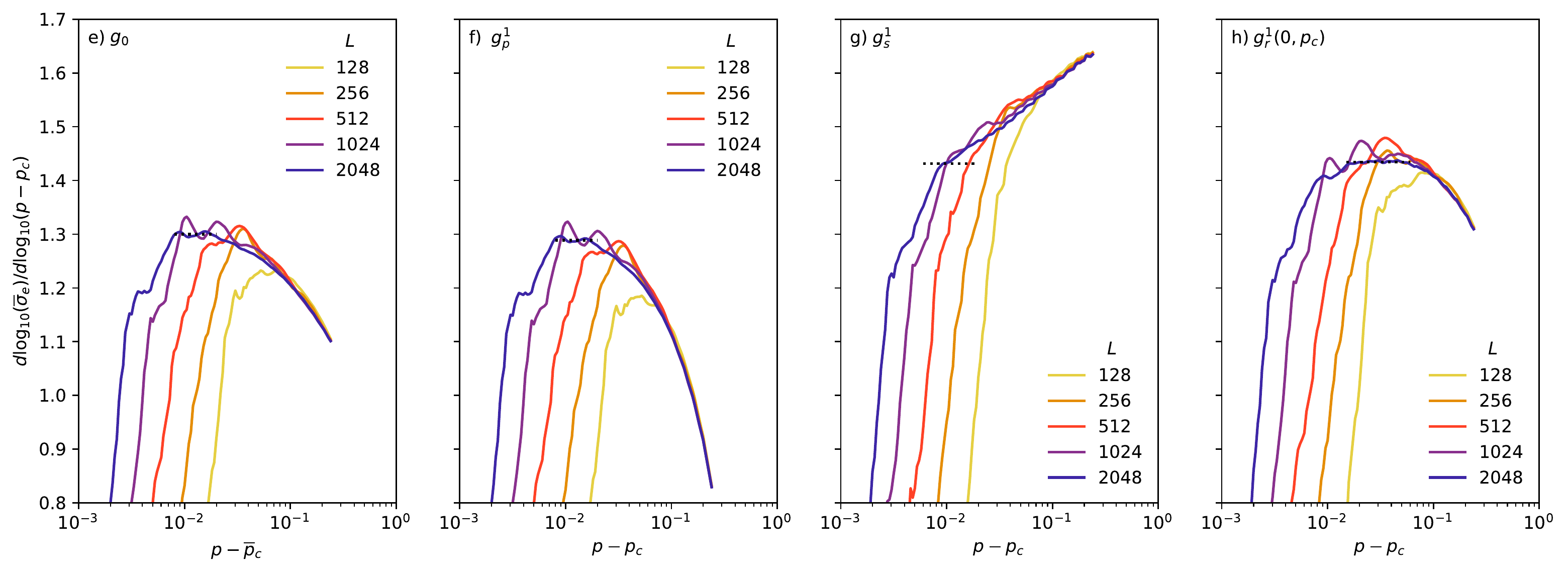}
\end{center}
\caption[]{Average effective conductivity $\effcond_e$ for 100 realizations for each 
conductance map $\condmap^1$ in (a)-(d), with the corresponding derivatives in 
(e)-(f). Note that for each of the 100 realizations we have used the same 
$p(e)$ distribution for the four different conductance maps. The slope is 
estimated in the range marked by the dotted line in the derivative plots, with 
the error estimates for the slopes being simply the difference between the 
minimal and maximal derivative value inside the given range. There is no 
plateau for the $\condmap_s^1$ model, and the dashed line range for model 
$\condmap_s^1$ in (g) was chosen to obtain a slope similar to the slope for 
model $\condmap_r^1$.}
\label{fig:combinedIntp}
\end{figure*}

\begin{figure*}[t]
\begin{center}
\includegraphics[width=2\columnwidth]{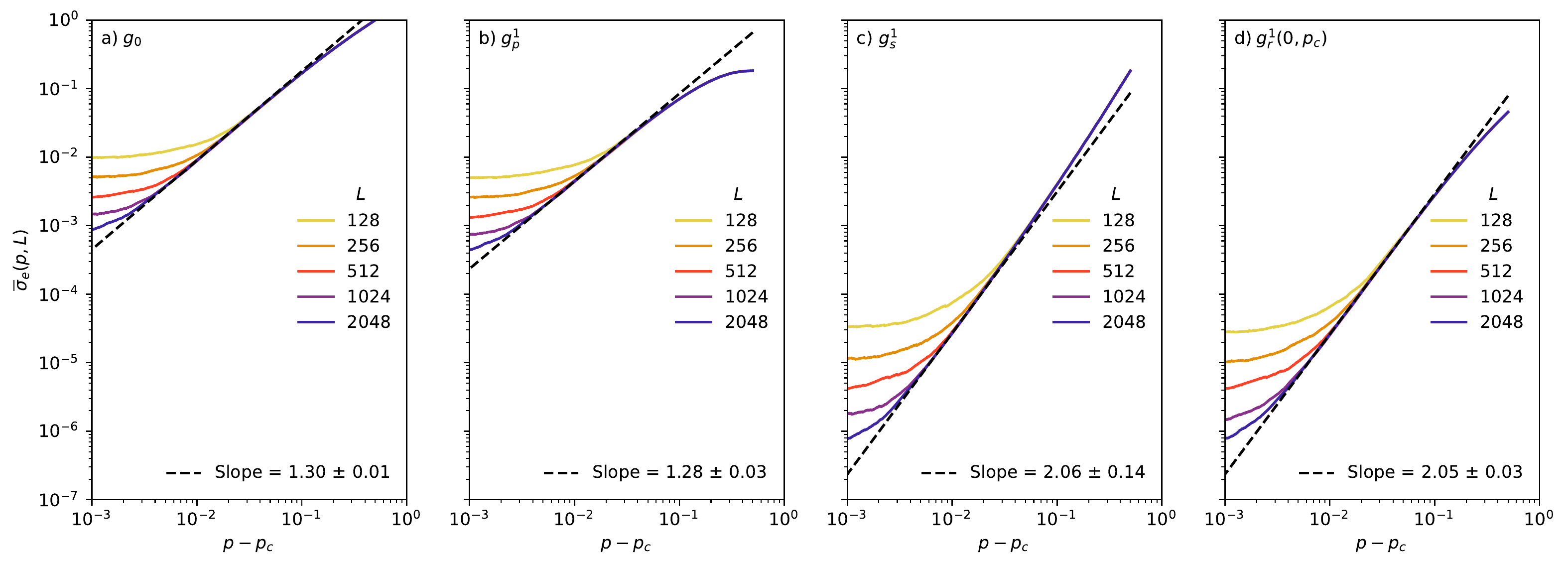}
\includegraphics[width=2\columnwidth]{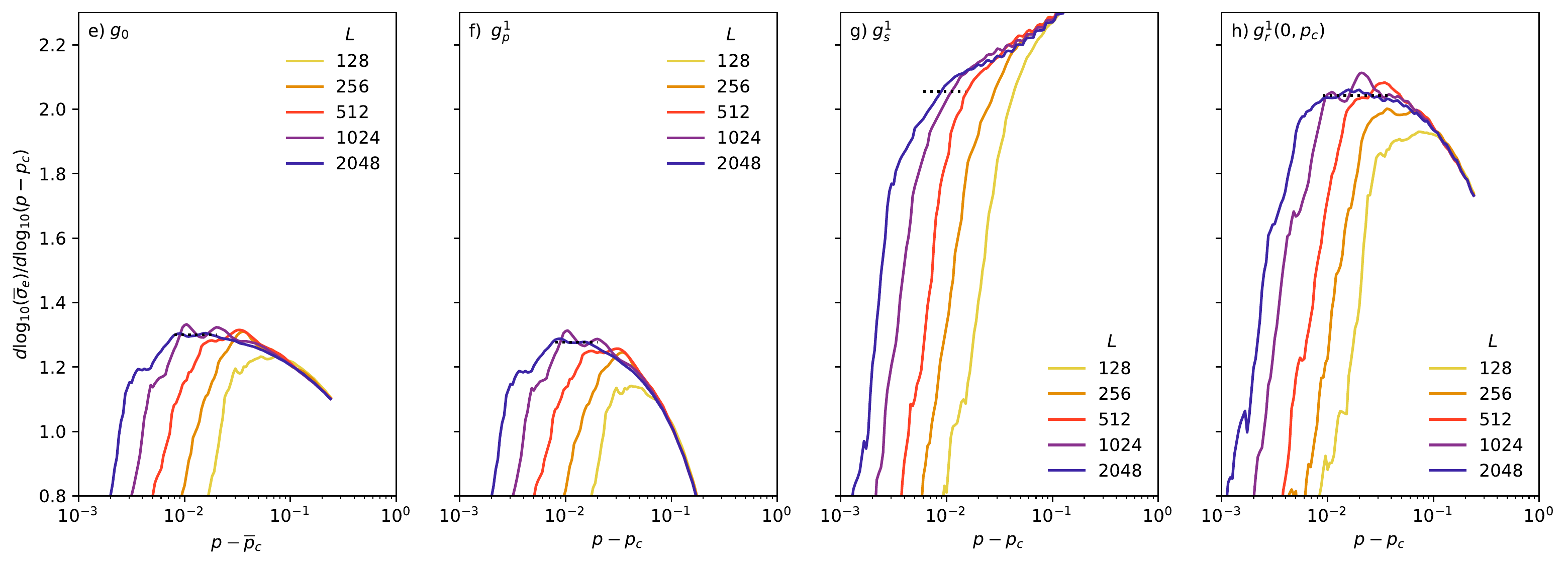}
\end{center}
\caption[]{Average effective conductivity $\effcond_e$ for 100 realizations for each 
conductance map $\condmap^2$ in (a)-(d), with the corresponding derivatives in 
(e) and (f). The slope is estimated in the range marked by the dotted line in 
the derivative plots, with the error estimates for the slopes being simply the 
difference between the minimal and maximal derivative value inside the given 
range. Note that the plots for $\condmap_o$, in (a) and (e), are equal to the 
corresponding ones in Figure \ref{fig:combinedIntp}; however, their $y$-scales 
differ. As with $\condmap_s^1$, there is no plateau for the $\condmap_s^2$ model, 
and the dashed line range for model $\condmap_s^2$ was chosen to obtain a slope 
similar to the slope for model $\condmap_r^2$.}
\label{fig:combinedIntp2}
\end{figure*}

As is well-known, near the percolation threshold $p_c$, the effective conductivity of the network $\condnet_o$ (i.e., the network resulting from the conductance map $\condmap_o$) follows the power law given in Eq.~\eqref{eq:condPowerLaw} with a critical exponent $t\simeq 1.3$. Figure 
\ref{fig:combinedIntp}(a) presents the dependence of the average of the effective conductivity $\effcond_e^o(p,L)$ of 100 realizations of the networks of type $\condnet_o(p)$, the standard percolation conductivity model, on both $L$, the 
linear size of the network, and $(p-p_c)$. Figure \ref{fig:combinedIntp}(e) 
shows the numerical derivatives of the curves in Fig.~\ref{fig:combinedIntp}
(a). We see that by increasing the size of the network the gradient reaches a 
plateau with a value close to 1.3 and, thus, $\condnet_o$ converge to a power 
law of type \eqref{eq:condPowerLaw} with a slope $t\simeq 1.3$, in agreement 
with the theoretical expectation.

Next, we investigate the critical exponent for the conductance model 
$\condmap_p^\tau$ by identifying an upper and lower bound for the exponent value.
The individual bond conductances of $\condnet_o$ are always larger or equal to 
the bond conductances of $\condnet_p^\tau$ for all $\tau\geq 0$, i.e., 
$\condmap_o\geq\condmap_p^\tau$ for $\tau\geq 0$. As a consequence of 
\citep[Lemma 11.4]{kesten_percolation_1982}, $\condmap_o\geq\condmap_p^\tau$ 
implies that $\effcond_e^o(p,L)\geq \effcond_e^p(p,L)$. If $\condnet_p^\tau$ follows a 
universal power law of type \eqref{eq:condPowerLaw} with exponent $t_p$, then 
$\effcond_e^p(p,L)\leq \effcond_e^o(p,L)$ implies that $t_p\geq t=1.3$. Thus, we have
identified a lower bound for the exponent $t_p$.

We now derive an upper bound for $t_p$. For all $p>0$, the smallest bond 
conductance value in $\condmap_p^\tau$ is $(1-p)^\tau$. If we let $p\condmap_o$ 
denote the network with all bond conductances equal to $p$, then 
$\condmap_p^\tau(p,L)>(1-p)^\tau\condmap_o(p,L)$ for all $p_c\leq p<1$. For each $\tau$, since 
$\effcond_e^p(1,L)>0$, there exists an $\epsilon>0$ such that $\condmap_p^\tau(p,L)>\epsilon
\condmap_o(p,L)$ for all $p_c\leq p\leq 1$. The effective conductivity of 
$\epsilon\condmap_o$ is $\epsilon \effcond_e^o$, where $\effcond_e^o$ is the effective 
conductivity of $\condmap_o$. As the effective conductivity of $\epsilon\condmap_o$
and $\condmap_o$ are equal up to a scaling with $\epsilon$, then $\epsilon 
\condmap_o$ has the same power law exponent in Eq.~\eqref{eq:condPowerLaw} as 
$\condmap_o$, namely, $t\simeq 1.3$. Using the same argument that was utilized 
for the lower bound, $\effcond_e^p(p,L)\geq\epsilon \effcond_e^o(p,L)$ implies that $t_p\leq t=1.3$. Since we then have the same lower and upper bound for $t_p$, namely,
$t\leq t_p\leq t$, we have $t_p=t\simeq 1.3$. Thus, networks of type 
$\condmap_p^\tau$ follow the traditional critical behavior when $p \to p_c$.

Next, we consider an alternative method for estimating $t_p$. As $p\to p_c$, 
the mass distribution of $\condmap_p^\tau$ will converge towards the distribution
$p_c-p(e)$, where $p(e)\in U(p_c,1)$. Thus, the mass distribution of 
$\condmap_p^\tau$ converges towards the mass distribution of a network of type 
$\condmap_r^\tau(p_c,1)$, i.e., a $\condmap_r^\tau$ function with $m(e)\in U(p_c,1)$. The networks $\condmap_r^\tau(p_c,1)$ and $\condmap_p^\tau$ are therefore 
expected to have the same properties when $p\to p_c$, including similar 
critical exponent (this will be substantiated further in the discussion on 
$\condmap_s$ below). We have conducted simulations to confirm such a convergence.

We now use $\condmap_r^\tau(p_c,1)$ to obtain the power law description for 
$\condmap_p^\tau$. The effective conductivity of $\condmap_r^\tau (p_c,1)$ is 
bounded from above by $\condmap_o$ and by $p_c^\tau \condmap_o$ from below. Since 
$p_c^\tau\condmap_o$ has the same critical exponent $t\simeq 1.3$ as $\condmap_o$, 
then, the critical exponent for $\condmap_r^\tau (p_c,1)$ is bounded from both 
above and below by $t\simeq 1.3$ and, thus, the exponent for $\condmap_r^\tau 
(p_c,1)$ is also $t\simeq 1.3$. As $\condmap_p^\tau$ and $\condmap_r^\tau(p_c,1)$ 
converge when $p \to p_c$, they have the same critical exponents, which 
provides an alternative proof that $t_p\simeq 1.3$.

The results for $t_p$ were verified by the simulations. Figure 
\ref{fig:combinedIntp}(b) and (f) present the average effective conductivity 
$\effcond_e^p(p,L)$ and its gradients for the model $\condmap_p^1$. Similarly, we show 
the average effective conductivity and gradients for $\condmap_p^2$ in Fig.~\ref{fig:combinedIntp2}(b) and (f). The inequality $\effcond_e^o(p,L)\geq \effcond_e^p(p,L)$ 
used to obtain the lower bound for $t_p$ can be verified by comparing Fig.~\ref{fig:combinedIntp}(a) with Fig.~\ref{fig:combinedIntp}(b) and Fig.
\ref{fig:combinedIntp2}(b). The slope of the $\effcond_e^p(p,L)$ curves, both for 
$\tau=1$ in Fig.~\ref{fig:combinedIntp}(f) and for $\tau=2$ in Fig.~\ref{fig:combinedIntp2}(f), converge towards a plateau. While the $\effcond_e^o$ and 
$\effcond_e^p$ curves have different heights, the plateaus of their gradients have 
similar heights. It is seen that the plateau values for $\condmap_p^1$ and 
$\condmap_p^2$ are in good agreement with the theoretical value of $1.3$. Note 
that the plots of the derivatives for the $\condmap_o$ and $\condmap_p^\tau$ models
have clear similarities, both for $\tau=1$ and 2, as we use the same $p(e)$ 
distribution for the $\condmap_o$ and $\condmap_p^\tau$ networks.

\begin{figure}[t]
\begin{center}
\includegraphics[width=0.9\columnwidth]{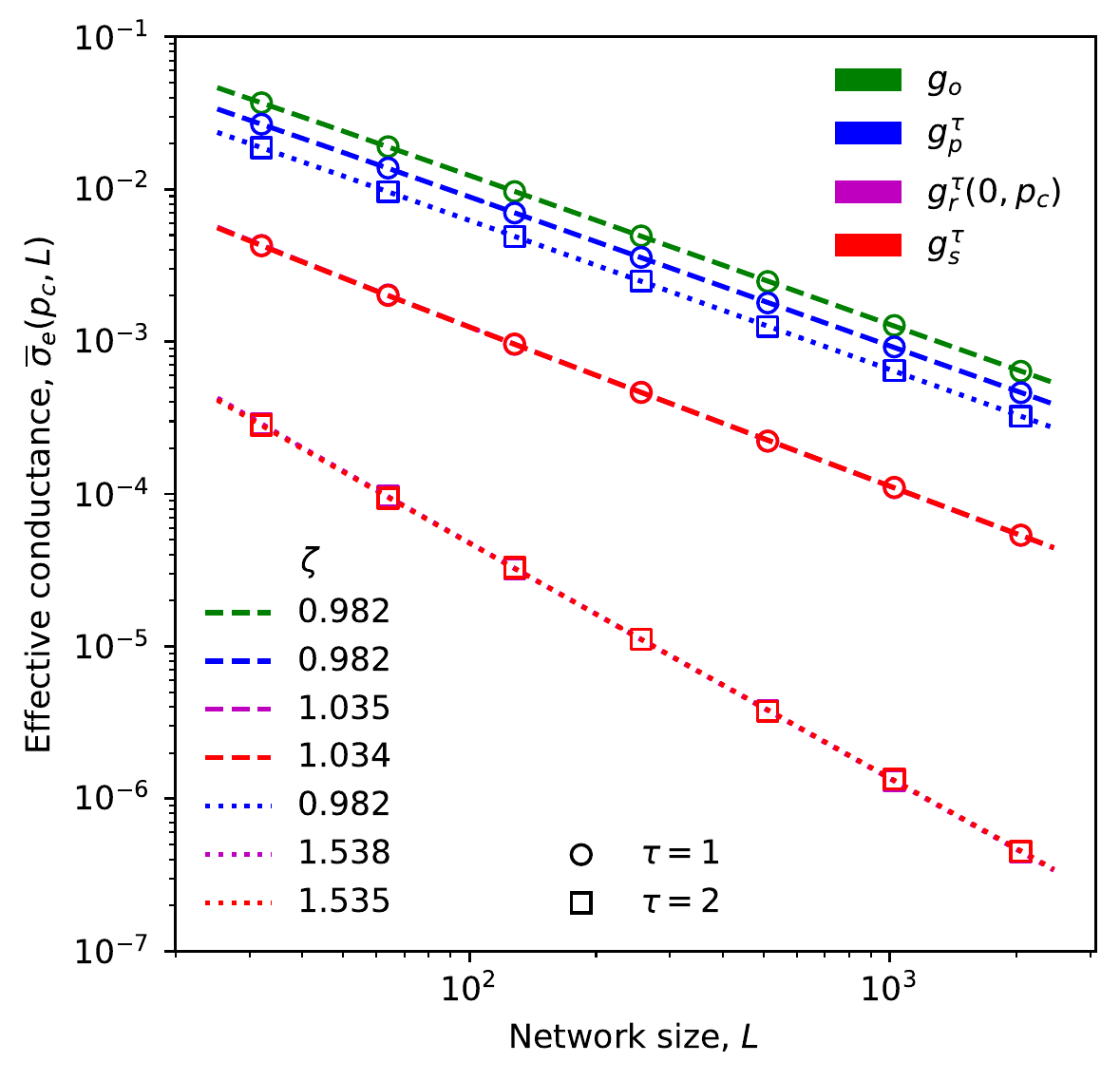}
\end{center}
\caption[]{Average effective conductivity $\effcond_e$ at the theoretical percolation 
threshold $p_c=0.5$ for more than 5000 realizations of each network size $L$. 
Note that the curve for $\condmap_r^\tau$ is covered by the curve for 
$\condmap_s^\tau$, as they are basically indistinguishable.}
\label{fig:finiteSizeScale}
\end{figure}

To further investigate the power laws for the various $\condmap$ functions, and in
particular $\condmap_p$, we consider finite-size scaling at $p_c$ \cite{sahimi_applications_2023,stauffer_introduction_2003}, namely, $\overline{g}_e(p_c)
\propto L^{-t/\nu}$, where $\overline{g}_e(p_c)$ is the average effective 
conductivity $\effcond_e(p_c)$ at the percolation threshold $p_c$ of a large number 
of network realizations, and $\nu$ is the critical exponent of percolation 
correlation length with $\nu=4/3$ in 2D. We tested linear regression using both
$L^{-\zeta}$ and curves with three free parameters of types suggested in 
\cite{sahimi_correction_1991}. The curve type yielding the best fit is of the 
form $L^{-\zeta}(a_1 - a_2/L)$, and is the plot type included in Fig.~\ref{fig:finiteSizeScale}. Note that the other curve types, including  
$L^{-\zeta}$, yielded similar $\zeta$-exponents.

Figure \ref{fig:finiteSizeScale} indicates that finite-size scaling yields an 
exponent of $\zeta=t/\nu\simeq 0.982$ for the standard percolation 
conductivity, corresponding to $\condmap_o$, close to the expected value of 
$t/\nu\simeq 0.975$. Note that $m_p^1>m_p^2$ since $1-p(e)<1$ (see Eq.~\ref{eq:evolveOrg}) and, thus, $\condmap_o>\condmap_p^1>\condmap_p^2$, as observed 
in Fig.~\ref{fig:finiteSizeScale}. As discussed above, we expect the same 
critical exponent for $\condmap_p^\tau$ as for $\condmap_o$. The models associated 
with $\condmap_p^\tau$ yield slopes similar to that of $\condmap_o$, and the 
computed $\zeta\approx 0.982$ are consistent with this expectation, yielding 
$t=\zeta\nu \simeq 1.31\simeq 1.3$. 

\begin{figure*}[t]
\begin{center}
a) \includegraphics[width=0.6\columnwidth]{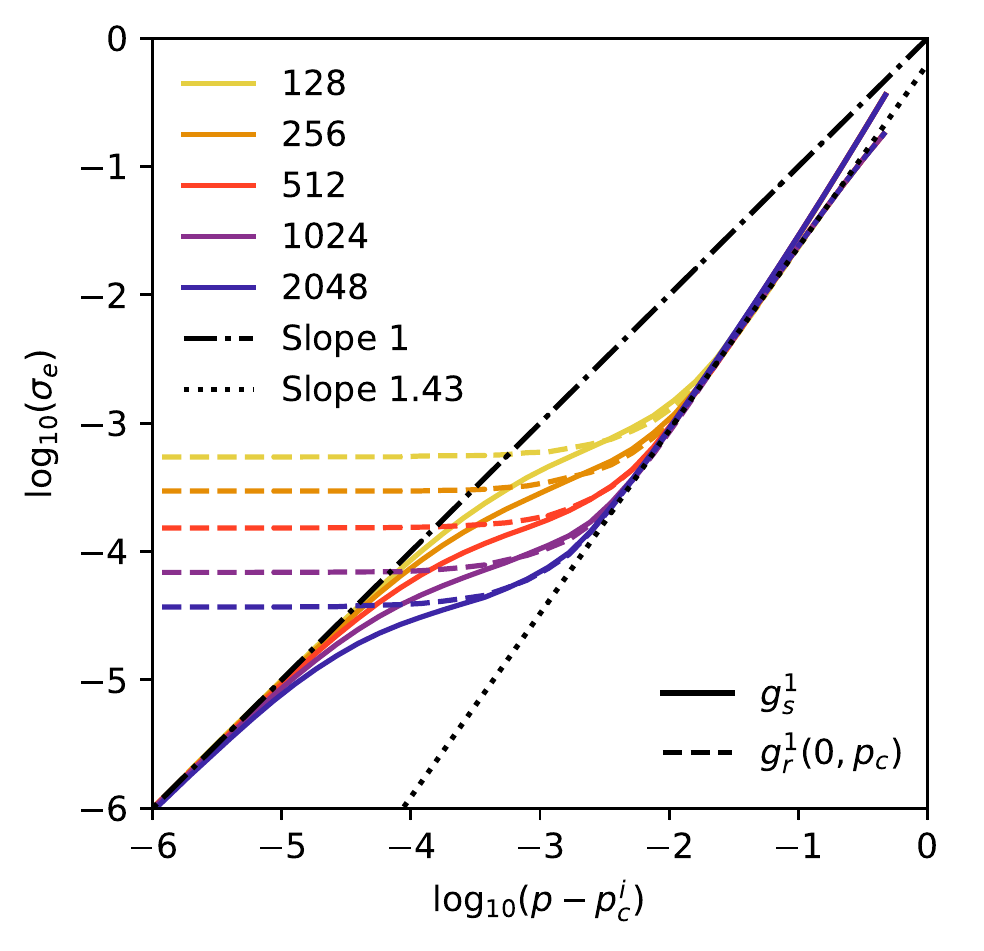}
b) \includegraphics[width=0.6\columnwidth]{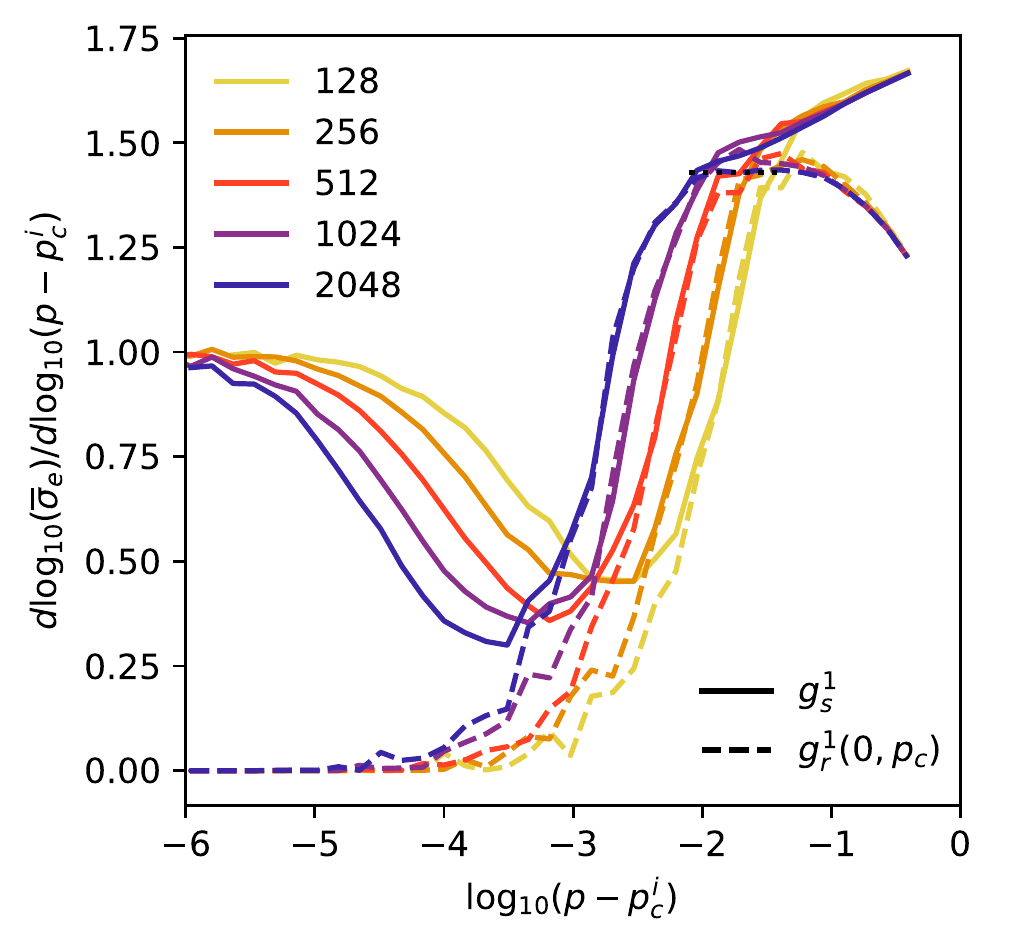}
c) \includegraphics[width=0.6\columnwidth]{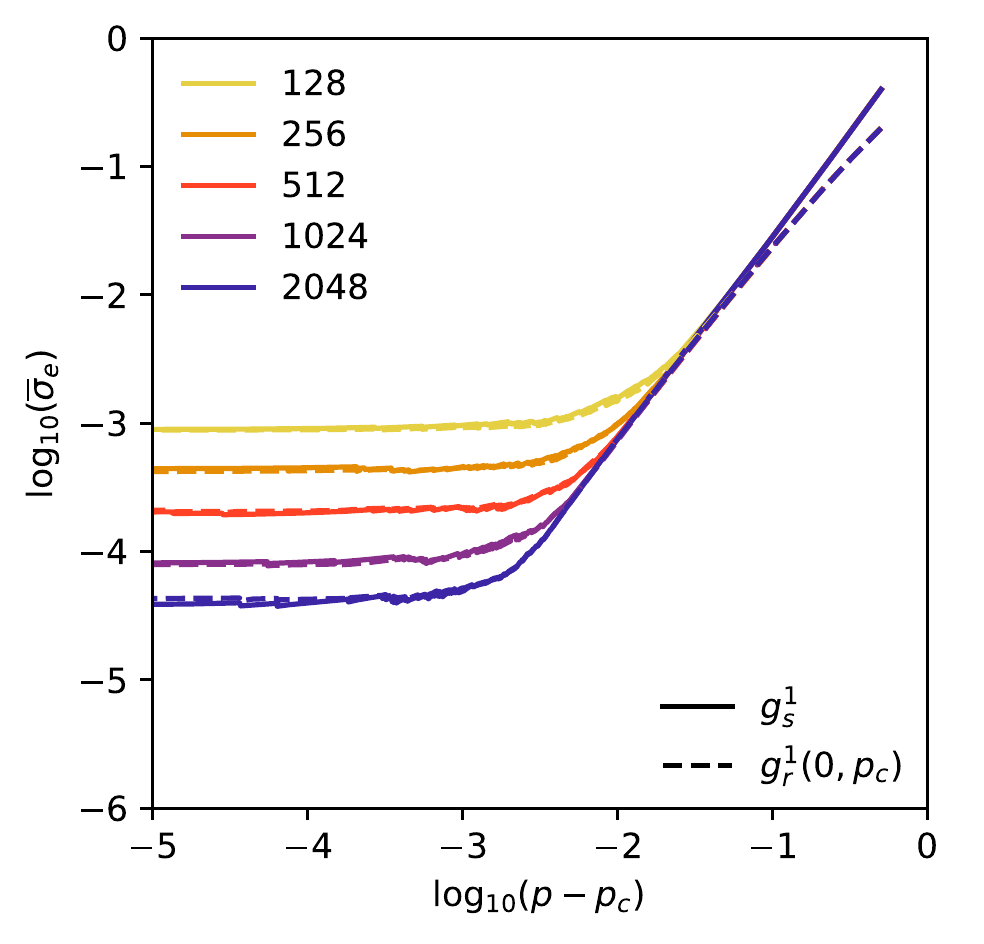}
\end{center}
\caption[]{(a) Average effective conductivity $\effcond_e$ for the same 100 
realizations as used in Fig.~\ref{fig:combinedIntp}(c) and (d). They are, 
however, plotted for the convergence towards their individual thresholds 
$p-p_c^i$. (b) Numerical derivatives of the curves in (a), with the dashed line
indicating the plateau of the $\condmap_r^1(p_c,1)$ curves. The plot in (c) is 
using the global percolation threshold $p_c$, instead of the individual 
percolation thresholds $p_c^i$.}
\label{fig:combinedThr}
\end{figure*}

\subsection{Conductance function $\condmap_s^\tau$}

A critical difference between models $\condmap_s$ and $\condmap_p$ is that the 
conductance distribution of the bonds in $\condmap_s$ diverge, which can cause 
non-universal behavior \cite{kogut_distribution-induced_1979,feng_transport_1987}. Conductance distributions and non-universal behavior will be discussed 
in the next section. As in the alternative derivation of $t_p$, we will use 
functions of type $\condmap_r^\tau$ to identify the critical exponents $t_s$ for 
$\condmap_s^\tau$.

Let $p_c^i$ be the individual percolation threshold for a given network (one realization of $p(e)$ values). The link with $p(e)=p_c^i$ is the bridging link,
$e_b$, which becomes a single connection that keeps the network connected when 
approaching the individual percolation threshold $p_c^i$. When $e_b$ is 
removed at $p=p_c^i$, the remaining network will be disconnected. The 
conductance of the bridging link will be $(p-p_c^i)^\tau\to 0$ when $p\to p_c^i$, whereas for all other links the conductance $(p-p(e))^\tau$ converges to a 
positive constant. Since the remainder of the network has finite conductance 
when $p\to p_c^i$, the resistance of the bridging link will dominate the 
resistance of the full network in the limit $p\to p_c^i$. Thus, the effective 
conductivity scales as $\effcond_e\propto (p-p_c^i)^\tau L^{2-d}$, when $p\to p_c^i$ 
for networks of spatial dimension $d$. In Fig.~\ref{fig:combinedThr}a) we 
present the effective conductivity of both $\condnet_s^1$ and 
$\condnet_r^1(0,p_c)$, indicating that the conductivity of $\condmap_s^\tau$ 
converges to the slope given by $\tau$, as expected from the derivation above. 

If we consider a two-dimensional network $\condnet'$ in which all other links 
than $e_b$ in $\condnet_s$ are replaced by superconductors, then the network 
$\condnet'$ will have a conductivity $\effcond_e'\propto (p-p_c^i)^\tau$ when $p\to 
p_c^i$. Thus, the development of the conductivity is of the power-law type Eq.~\eqref{eq:condPowerLaw} with critical exponent $\tau$. Since the 
effective conductivity of $\condnet'$ is larger than the conductivity of 
$\condnet_s^\tau$, i.e., $\effcond_e'> \effcond_e^s$, we see that the critical exponent $t_s$ 
must be bounded below as $t_s\geq\tau$. Note that, as the conductivity of 
$\condmap_s^\tau$ is always smaller than the conductivity of $\condmap_o$ when 
$\tau>0$, $\effcond_e^o > \effcond_e^s$, we also have $t_s\geq t=1.3$. Thus, in general, we 
have, $t_s\geq \max(t,\tau)$, giving a lower bound for $t_s$.

Consider the situation in which $L\gg\xi$, i.e., one in which $L$ is large 
compared to the correlation length $\xi$ of percolation. In this limit there 
are no singly-connected bonds; according to \cite{stauffer_introduction_2003} 
the minimum cut contains approximately $L/\xi$ bonds. As the network is well 
connected when $L\gg\xi$, we can disregard the effect of the conductance of 
$e_b$ vanishing when $p\to p_c^i$, as $e_b$ is then on one of many connected 
paths in the infinite percolation cluster. The network will have a mass
distribution equivalent to that in $\condmap_r^\tau(0,p_c^i)$ when $p\to p_c^i$. 
To compare our network to $\condmap_r^\tau(0,p_c)$, we need $p\simeq p_c$ for the
distribution of bond conductances in $\condmap_s^\tau$ to be similar to that in
$\condmap_r^\tau(0,p_c)$. This requirement does not, however, scale with $L$, so 
that we can expect the two conductance distributions $\condmap_s^\tau$ and 
$\condmap_r^\tau(0,p_c)$ to converge at the same values of $p$, independent of 
the size $L$. Therefore, for large $L$ we can expect a region of $p$ values 
where $\condmap_s^\tau \simeq \condmap_r^\tau(0,p_c)$, i.e., where $L\gg\xi$ and $p
\simeq p_c$.

In Fig.~\ref{fig:combinedThr} we present the results for both $\condmap_s^1$ and 
$\condmap_r^1(0,p_c)$. As seen in the figure, $\condmap_s^1$ and $\condmap_r^1(0,p_c)
$ differ for both large and small values of $p-p_c^i$; they are, however, 
similar for a range of intermediate values that correspond to the region in
which $L\gg\xi$ and $p\simeq p_c$. We also observe that the two curves diverge 
when $p\to p_c^i$: In this case, we have $L\ll\xi$ and, thus, the link $e_b$ 
will become the single bridging link. Since the weight $\condmap_s^\tau(p,e_b)\to
0$ when $p\to p_c^i$, this conductance will begin dominating the overall 
conductance of the network as described above, and the conductance will vanish 
by the power law, $\effcond_e^s \propto (p - p_c^i)^\tau$, as $p \to p_c^i$. This is 
in contrast to the $\condmap_r^\tau$ network, for which the bridging link $e_b$ 
has a finite conductance, $\condmap_r^\tau(p,e_b)>0$ and, thus, $\effcond_e^r$ 
converges to a finite value when $p\to p_c^i$. The two conductance descriptions
$\condmap_s^\tau$ and $\condmap_r^\tau$ must, therefore, begin to diverge when $p\to p_c^i$, and Fig.~\ref{fig:combinedThr} indicates that they do. 

While the conductivities have clearly different trajectories when plotted 
versus their individual percolation thresholds $p_c^i$, the difference becomes 
insignificant when one uses instead the traditional averaging $p-p_c$, where 
$p_c$ is the percolation threshold for an infinite network. Let $p_{av}= 
\left< p_c^i \right>$ be the average of the percolation thresholds for the 
individual networks, and let $\Delta=\sqrt{\left<(p_c^i)^2\right>-\left<p_c^i
\right>^2}$ be the standard deviation of the individual percolation thresholds.
The two values are known to scale as $p_{av}-p_c\propto L^{-1/\nu}$ and 
$\Delta\propto L^{-1/\nu}$ \cite[p. 73]{stauffer_introduction_2003}. The 
standard deviation of the individual percolation thresholds $\Delta$ is larger 
than the difference between $p_{av}$ and $p_c$; thus, the $\Delta \propto 
L^{-1/\nu}$ correspondence will be of importance to us. The difference between 
the $\condmap_s^\tau$ and $\condmap_r^\tau(0,p_c)$ models when $p\to p_c^i$ is 
expected to be reflected in the $p-p_c$ curves only if $\Delta$ is smaller than
the onset of divergence between the $\condmap_s$ and $\condmap_r$ curves. In Fig.~\ref{fig:combinedThr}(c) we have plotted the results for $p-p_c$. There is no 
evident difference between the curves, indicating that $\Delta$ is larger than
the onset of the divergence observed in Fig.~\ref{fig:combinedThr}(a) and (b). 

Based on the above derivations, the power laws for $\condmap_s$ and $\condmap_r$ 
are expected to be the same, and should be bounded from below by 
$\max(t,\tau)$. This is corroborated by the results in Fig.~\ref{fig:finiteSizeScale}, where the results for $\condmap_r^\tau(0,p_c)$ and 
$\condmap_s^\tau$ are almost identical for both values of $\tau$. For $\tau=1$ 
they indicate $\zeta=t_s/\nu\simeq 1.034$, which yields a non-universal scaling
exponent of $t_s\simeq 1.38\geq t=\max(t,\tau)$. For $\tau= 2$ we have $\zeta 
\simeq 1.535$, yielding $t_s\simeq 2.05\geq\tau=\max(t, \tau)$. 

The results for $\condmap_s^1$ are presented in Figs.~\ref{fig:combinedIntp}(c) 
and (g), and those for $\condmap_s^2$ are shown in Figs.~\ref{fig:combinedIntp2}
(c) and (g). Since $\condmap_o>\condmap_p^\tau>\condmap_r^\tau$, we have $\effcond_e^o>\effcond_e^p>\effcond_e^r(0,p_c)$. It is evident from Fig.~\ref{fig:combinedIntp}g) that even the 
largest network size, $L=2048$, does not produce a plateau for the gradient. 
We thus plot $\condmap_r^1(0,p_c)$ in Figs.~\ref{fig:combinedIntp}(d) and (h). 
The derivative indicates a plateau, however, at a value around $t_s\simeq 
1.43$. This is higher than, $t_s=1.38$, obtained from the finite-size scaling 
above. For $\tau=2$, as seen in Figs.~\ref{fig:combinedIntp}(d) and (h), we 
obtain a slope of $t_s\simeq 2.05$, which is in agreement with the finite-size 
scaling above. These results will be discussed further in the next section.

\section{Discussion}

In the previous section, we investigated the power laws for the effective
conductivity of evolving networks, $\condmap_p^\tau$ and $\condmap_s^\tau$, 
introduced in this paper. We argued that the effective conductivities of these 
networks follow the same power laws as the networks $\condmap_r^\tau(p_c,1)$ and 
$\condmap_r^\tau(0,p_c)$, respectively. 

\begin{figure}[t]
\begin{center}
\includegraphics[width=0.8\columnwidth]{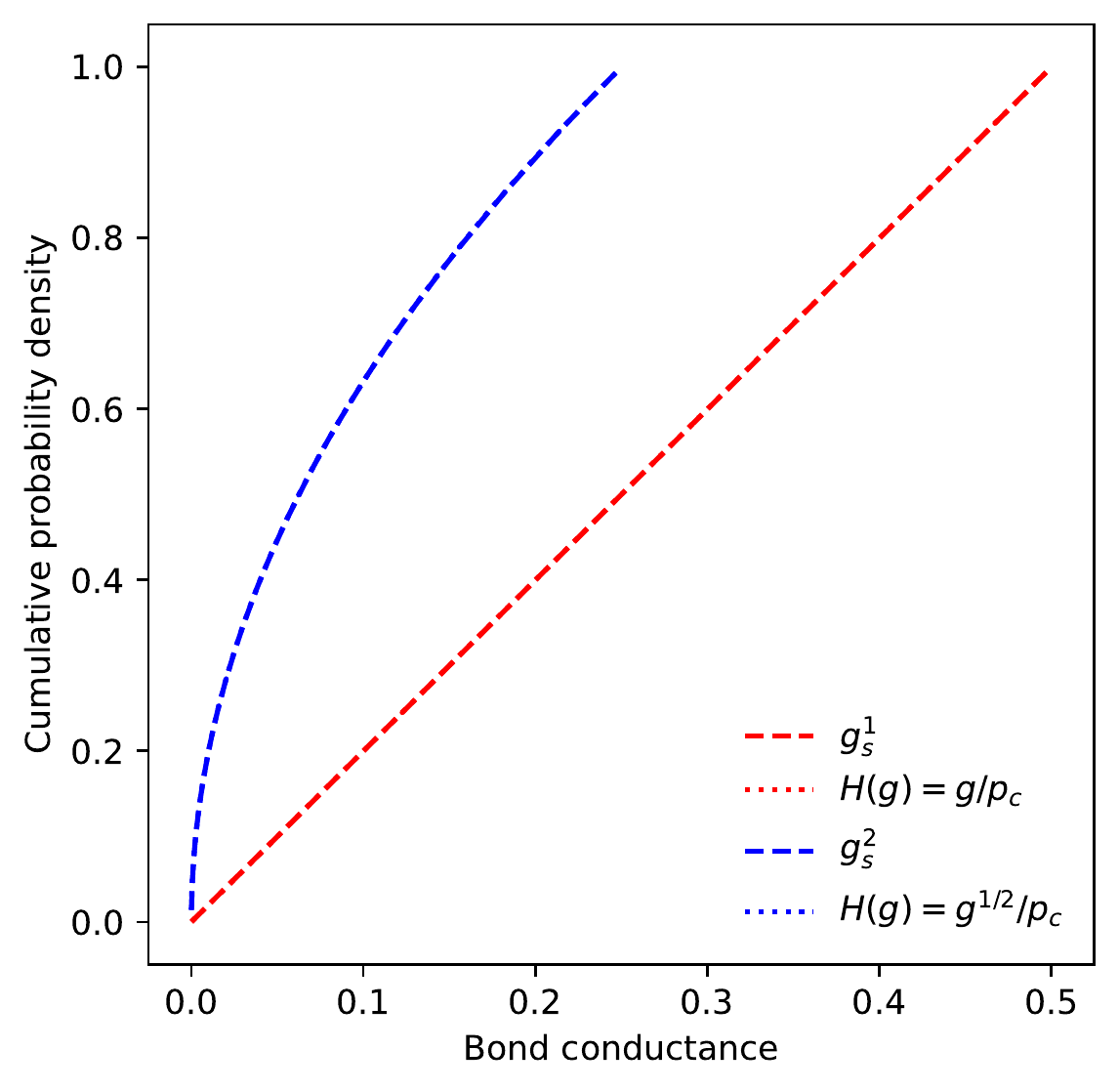}
\end{center}
\caption[]{Cumulative conductance distribution for $\condmap_s^\tau$ for 100 
realizations of size $L=512$, together with the functional relationships 
describing the distribution for $\condmap_r^\tau(0,p_c)$. The functional 
relationships are covered by distributions for $\condmap_s^\tau$}
\label{fig:cumDist}
\end{figure}

Non-universality has been observed for networks whose distribution of bond 
conductances diverges when the conductance values go to zero
\cite{kogut_distribution-induced_1979,feng_transport_1987}. For $\condmap_r^\tau (0,p_c)$ we have a uniform distribution of bond mass values in the range 
$[0,p_c]$, and the conductance for a bond of mass $m$ is $\bondcond = m^\tau$. 
The probability of having a mass smaller than $m$ is $m/p_c$. Thus, the 
probability of having a conductance smaller than $\bondcond=m^\tau$ becomes 
$m/p_c=\bondcond^{1/\tau}/p_c$, and the cumulative conductance distribution is 
given by
\begin{equation}
  H(\bondcond) = \bondcond^{1/\tau}/p_c \quad ,
  \label{eq:cumDist}
\end{equation}
for $\bondcond \in (0,p_c^\tau)$. In Fig.~\ref{fig:cumDist} we present the 
conductance distribution in $\condmap_s^\tau$ for the backbone at $p=p_c$, 
together with the distribution function in Eq.~\eqref{eq:cumDist}. We observe 
an equivalent distribution for $\condmap_s^\tau$ as $\condmap_r^\tau (0,p_c)$.

If we scale the conductances in the range $(0,p_c^\tau=2^{-\tau})$ to the 
range $(0,1)$ (with the above notation, we, thus, consider $p_c^{-\tau} \condmap_r^\tau(0,p_c)$), we have the cumulative probability $H(\bondcond)=\bondcond^{1/\tau}$, which yields  the probability distribution 
\begin{equation}
h(\bondcond)=\frac{1}{\tau}\bondcond^{1/\tau-1}=(1-\alpha)\bondcond^{-\alpha} \quad ,
\label{eq:condDist}
\end{equation}
where the last term is on the form used in \cite{kogut_distribution-induced_1979}, obtained from $\alpha=1-1/\tau$. For $\tau>1$ we have a negative 
exponent for $\bondcond$ in Eq.~\eqref{eq:condDist}, making $h(\bondcond)$ diverge 
when the conductance $\bondcond\to 0$. According to \cite{kogut_distribution-induced_1979}, we then have $\effcond_e\propto(p-p_c)^{t_r}$, where 
$t_r=t+\alpha/(1-\alpha)=t+\tau-1$, with $t$ being the standard conductivity 
exponent, with $t\simeq 1.3$ for two-dimensional networks, as mentioned above. 
Note also that other authors reported different values for $t_r$, with $0<t_r-
t < 3/2$ for $\tau=2$ according to \cite{feng_transport_1987}. In 
\cite{straley_non-universal_1982} the non-universal exponent is given as 
$t_r=\max(t,(1-\alpha)^{-1})=\max(t,\tau)$, which is exactly the lower bound 
we obtained for $\condmap_s$ above.

For $\tau=1$ the literature indicates that for the $\condmap_r^\tau$ model $t_r= 
t\simeq 1.3$. Our derivations above should have yielded $t_s=t_r=t$, but our 
numerically computed values for $t_s$ are higher than this, with $t_s\simeq 1.38$ by finite-size scaling and $t_s\simeq 1.43$ through investigating the 
gradient of the curves $\condmap_s(L,p)$. It has been reported that the 
universality constant for $t_r$ is difficult to obtain as logarithmic 
corrections set in for $\tau=1$ \cite{straley_non-universal_1982}. Our computed
values are, however, in excellent agreement with estimates from similar 
numerical simulations for the $\condmap_r^\tau$ model 
\cite{flukiger_nonuniversal_2008}.

For $\tau=2$, the literature differs on the value of $t_r$, with $1.3< t_r<
2.8$; according to \cite{feng_transport_1987}, $t_r\simeq 2.3$; according to 
\cite{kogut_distribution-induced_1979}, and $t_r=2$ according to 
\cite{straley_non-universal_1982}. Our estimate of $t_s\simeq 2.05$ is within 
the spread of the $t_r$ values for the $\condmap_r^\tau$ model, as indicated by 
the aforementioned authors.

\section{Summary}

We introduced two types of evolving networks that are related to natural and 
industrial processes, such as clogging, precipitation, and dissolution. One 
model, $\condmap_p^\tau$, represents clogging processes that tend to block the 
lowest conducting bonds. The second model, $\condmap_s^\tau$, represents 
precipitation processes that reduce the conductance of all bonds similarly. 
The mass distribution is linked to the conductance by the exponent $\tau$, 
where $\tau=1$ represents electrical conductance or diffusion, while $\tau=2$ 
represents fluid flow.

The effective conductivity of the models that we introduced behaves differently 
from that of the traditional networks $\condmap_o$ with constant bond 
conductance. We showed, however, that the power laws $\effcond_e^p\propto 
(p-p_c)^{t_p}$ for $\condmap_p^\tau$ still belong to the standard universality 
class with exponent $t_p=t\simeq 1.3$.

The effective conductivity of the $\condmap_s^\tau$ model follows a power law 
similar to $\condmap_r^\tau(0,p_c)$. The effective conductivity of the 
$\condmap_r^\tau(0,p_c)$ model is known in the literature to have non-universal 
power laws near the percolation threshold, and we have the same 
non-universality for $\condmap_s^\tau$. The conductivity of the $\condmap_s^\tau$ 
model has, however, a radically different behavior than $\condmap_r^\tau(0,p_c)$,
when we consider convergence towards individual percolation thresholds, $p\to 
p_c^i$. In this limit the $\condmap_s^\tau$ conductivity scales as $\effcond_e^s\propto 
(p - p_c^i)^\tau$, which leads to a lower bound $t_s\geq\max(t,\tau)$ for the
power law, $\effcond_e^s \propto (p-p_c)^{t_s}$. As the effective conductivity of 
both $\condmap_s$ and $\condmap_r(0,p_c)$ follow the same power laws, this yields 
the same lower bound for $\condmap_r^\tau(0,p_c)$, namely, the lower bound 
$t_r\geq\max(t,\tau)$.

\section*{Acknowledgments}

The first author is supported by the Research Council of Norway (Centers of 
Excellence funding scheme, project number 262644, PoreLab). The second author 
is grateful to the National Science Foundation for partial support of his work 
through grant CBET 2000966.

\bibliography{references}

\end{document}